\documentclass[article]{JHEP3}

\usepackage{epsfig,multirow}
\usepackage{amsmath}
\usepackage{graphicx}
\usepackage[latin1]{inputenc}
\usepackage{amssymb}
\usepackage{amsmath}
\usepackage{verbatim}
\usepackage{epsfig}

\newcommand{\GeV}{\mathrm{\,GeV}}

\newcommand{\ie}{{\it i.e.~}}
\newcommand{\eg}{{\it e.g.~}}

\newcommand{\pythia}{{\sc Pythia~}}
\newcommand{\bridge}{{\sc BRIDGE~}}
\newcommand{\herwig}{{\sc Herwig~}}
\newcommand{\herwigb}{{\sc Herwig}}

\newcommand{\madgraph}{{\sc MadGraph5~}}
\newcommand{\madgraphb}{{\sc MadGraph5}}
\newcommand{\MCFM}{{\sc MCFM~}}
\newcommand{\MCFMb}{{\sc MCFM}}
\newcommand{\madevent}{{\sc MadEvent~}}
\newcommand{\madeventb}{{\sc MadEvent}}
\newcommand{\madspin}{{\sc MadSpin~}}
\newcommand{\madspinb}{{\sc MadSpin}}
\newcommand{\powheg}{{\sc Powheg~}}

\newcommand{\mcatnlo}{{MC@NLO~}}
\newcommand{\amcatnlo}{{aMC@NLO~}}
\newcommand{\amcatnlob}{{aMC@NLO}}

\preprint{
 NIKHEF/2012-021 \\
CERN-PH-TH/2012-329}
\title{
Automatic spin-entangled decays of heavy resonances in Monte Carlo simulations} 
\author{Pierre Artoisenet\\
  Nikhef Theory Group, Science Park 105,  1098 XG Amsterdam, The Netherlands \\
  E-mail: \email{partois@nikhef.nl}}
\author{Rikkert Frederix\\
  PH Department, TH Unit, CERN, CH-1211 Geneva 23, Switzerland \\
  E-mail: \email{rikkert.frederix@cern.ch}}
\author{Olivier Mattelaer \\
  Department of Physics, University of Illinois at Urbana-Champaign, Urbana, IL 61801. \\
  E-mail: \email{omatt@illinois.edu}}
\author{Robbert Rietkerk\\
  Institute for Theoretical Physics, University of Amsterdam,  Science Park 904, 1090 GL Amsterdam, The Netherlands and \\
  Nikhef Theory Group, Science Park 105,  1098 XG Amsterdam, The Netherlands \\
  E-mail: \email{robbertr@nikhef.nl}}
\abstract{
We present a general method that allows one to decay narrow resonances 
in Les Houches Monte Carlo events in an efficient and accurate way. The procedure preserves both spin
correlation and finite width effects to a very good accuracy, and 
is therefore particularly suited for the decay of resonances in production 
events generated at next-to-leading-order accuracy. The method is implemented as a generic tool in  the \madgraph    
framework, giving access to a very large set
of possible applications. We illustrate the validity of the method and the code by applying it to the case of single top and top quark pair production, and show its capabilities on the case of top quark pair production in association with a Higgs boson.
}


\begin{document}

\section{Introduction}

The analysis of scattering processes involving heavy resonances 
at the Large Hadron Collider (LHC) is of essential 
importance. One of the reasons is that the Higgs boson
couples primarily to heavy particles such as the 
top quark and the weak bosons. Also many
theories beyond the Standard Model predict the existence 
of unstable particles at the TeV scale. 
These unstable particles are not observed directly, but
they initiate a cascade of decays (also called a \textit{decay branch})
ending up with particles that may be observed in the detector.
Such processes lead to a rich 
phenomenology, because the spin of the heavy resonances
and the type of coupling to other fields 
imply non-trivial angular correlations among the final-state particles
inside a given decay branch (a phenomenon also called 
\textit{decay spin correlation effects}) or among the particles 
from distinct decay branches (a phenomenon also called  
\textit{production spin correlation effects}).

Nowadays virtually all experimental analyses at the LHC
rely on tools to simulate scattering events and 
their reconstruction in the detector.  
As Monte Carlo generators have become an increasingly 
important tool for collider physics, a large effort has been 
devoted to improving their accuracy, in particular, by 
developing parton-level generators based on next-to-leading-order (NLO) QCD matrix elements.
In this respect, the simulation of production events with heavy resonances 
at next-to-leading-order accuracy in QCD is particularly challenging,
because the decay pattern of these resonances may lead to signatures 
characterized by a large particle multiplicity in the final state.
Because of the intrinsic complexity of next-to-leading-order amplitudes for large multiplicities of external particles,
the efficiency of the current algorithms  
imposes some limitations on the range of 
possible applications. 
In many instances, some approximations must be made,
otherwise the generation of events cannot be handled in practice. 

In this context, one well-known simplification is the 
narrow width approximation, which delivers a good accuracy
in the case  of a resonance with a
  width $\Gamma$ much smaller than its mass. 
In this approximation, intermediate resonances 
are put on their mass shell, 
significantly simplifying the structure of  QCD corrections
since radiative corrections in the production and in the decay
do not interfere in the limit $\Gamma \rightarrow 0 $.
As a result, QCD corrections in the production and in the decay
 can be handled separately. 
This scheme is implemented in 
Monte Carlo integrators such  as
 \MCFM\cite{Campbell:2004ch,Campbell:2005bb,Badger:2010mg} 
 for specific processes
involving the top quark. In that implementation, spin
correlation effects are retained at next-to-leading-order
accuracy (in the limit $\Gamma \rightarrow 0 $).

However, as far as Monte Carlo event generators
are concerned, the previous approach may not be the 
optimal one. Indeed, in the case of processes with 
complicated decay patterns, the efficiency of the generation
of unweighted events (\ie events with the same weight) 
becomes a serious issue.
This problem of efficiency may be overcome by considering 
a stronger assumption, sometimes called the 
 \textit{decay chain} approximation~\cite{Richardson:2001df} 
which consists of factorizing 
 the squared amplitude into a 
production factor and a decay factor. Using this scheme,
the generation of unweighted events can be split into
(1) the generation of undecayed events in
which these resonances appear as on-shell final state particles and
(2) the decay of the resonances in each production events.
Off-shell effects may partly be recovered by smearing 
the invariant mass of each heavy resonance 
according to a Breit-Wigner distribution, and 
by reshuffling the other momenta in the undecayed event.
In the simplest implementation of this scheme, 
each resonance is decayed according to a uniform
distribution in its rest frame, in which 
case spin correlation effects are lost.
To gain accuracy, kinematics of the decay 
products can be generated randomly according 
to the squared amplitudes for the decay processes~\cite{Meade:2007js}.
In this latter case, decay spin correlation effects are partially
preserved, while production spin correlation effects
are lost as a result of the 
production vs. decay factorization 
at the squared amplitude level.

In the search for the optimal approach 
to generate events with intermediate resonances, 
a compromise must be reached between the efficiency and the 
accuracy of the approach. In this spirit, 
Frixione, Laenen, Motylinski and Webber~\cite{Frixione:2007zp}
have proposed a convenient procedure (called the FLMW procedure 
in this paper) for Monte Carlo generators  which not only preserves
the efficiency inherent to the decay chain approximation, 
but also includes nearly all spin correlation effects
at next-to-leading-order accuracy. 
They implemented this scheme in \mcatnlo\cite{Frixione:2002ik,Frixione:2005vw} 
and in \powheg\cite{Frixione:2007nw}
for specific processes involving the production of top quarks and weak bosons.
The accuracy of the scheme was demonstrated for these processes.

Since the work of Frixione \textit{et al}, there has been 
a tremendous progress in automating next-to-leading-order calculations.
A new milestone has been 
reached by promoting Monte Carlo generators into the era of 
fully automated NLO event generators~\cite{Alioli:2010xd,Frederix:2011zi,Hoeche:2011fd}. This breakthrough has 
opened various interesting perspectives on hadron collider
phenomenology, by allowing the simulation of a new class
of processes at next-to-leading-order accuracy. Among 
these processes, the production of heavy resonances 
--such as top quark pair production in association with 
a Higgs boson~\cite{Frederix:2011zi,Garzelli:2011vp}-- are directly relevant for the ongoing phenomenology 
at the Large Hadron Collider. 
Even though these automated NLO Monte Carlo generators feature, in principle, no restrictions on
complexity of the process and particle multiplicity, in practice the CPU cost becomes enormous
for high-multiplicity final states. Most of the current tools cannot simulate the full production and decay at
NLO accuracy in a reasonable amount of time; only the 
generation of undecayed events at next-to-leading order
is feasible. 

Some  frameworks already exist to decay heavy resonances 
in undecayed events. However the existing tools
are either limited to specific processes (\eg the implementation
in \mcatnlo\cite{Frixione:2007zp}), or they do not 
provide an acceptable accuracy given the current state-of-the-art
of simulation techniques (\eg the program \bridge\cite{Meade:2007js}
or
 the generic decay routines in \pythia\cite{Sjostrand:2006za,Sjostrand:2007gs}
or \herwig\cite{Corcella:2000bw}). 
In this work, we provide a generic and accurate algorithm
to decay heavy resonances in undecayed events generated at next-to-leading-order accuracy. We revisit the FLMW  procedure
and we demonstrate that this procedure can be fully 
automated. Special care is taken to handle 
 the off-shell effects and the reshuffling 
of the momenta.
We implement this algorithm in the \madgraph framework~\cite{Alwall:2011uj} and
we dub this new tool \madspinb.

\section{Spin correlations in the decay}
We start by introducing some notation.
The kinematics $X$ of an event $E_X$ can be parametrized by 
a set of independent variables $\textbf{x}=x^1, \dots , 
x^r$ each of them in the range $[0,1]$. The  function $\phi: \textbf{x} \rightarrow X$ that maps the variables $\textbf{x}$ onto the kinematics $X$ is usually called phase-space mapping.
In the corresponding decayed event $E_{X_\textrm{dk}}$ (with corresponding kinematics $X_{\textrm{dk}}$)
an additional set of independent variables 
$\textbf{y}=y^1, \dots , y^s$ is needed to parametrize
the kinematics of the decay products. In this work, 
the variables $\textbf{y}$ for each resonance decay are 
always in one-to-one correspondence with the invariant mass of the resonance
 and the angles defining the direction of the decay products 
in the rest frame of the resonance.   

Next-to-leading-order generators typically produce 
a list of events, where each event $E_X$ is characterized by the fully 
exclusive kinematics $X$.
The FLMW procedure~\cite{Frixione:2007zp}
allows one to generate the decays of narrow resonances in the event $E_X$
according to distributions that retain both decay and production spin correlation
effects.
For each undecayed event,
the procedure goes as follows.
\begin{enumerate}
\item The variables $\textbf{y}$ characterizing the decay 
 of the resonances are generated randomly,
     considering --for each decay-- a uniform distribution of the decay products in the rest frame of the decaying particle,
and the decayed event with kinematics $X_\textrm{dk}(\textbf{x},\textbf{y})$ built upon these variables
is reconstructed.

\item The differential cross section 
       $d\sigma / d\textbf{x}d\textbf{y}$
 associated with the decayed process is evaluated with tree-level matrix elements at the 
 decayed event with kinematics $X_\textrm{dk}(\textbf{x},\textbf{y})$ 
      and an unweighting procedure with respect to 
      the maximum weight is used to decide whether this 
      decayed event should be kept or not.
      More precisely, if 
      $W_{\textrm{max}}(X)$ represents the maximum value
      of the differential cross section  
       over the range
      of possible decay 
      configurations $\textbf{y}$'s, a random number $r$ is 
      generated uniformly between 0 and 1,
      and the decayed event $X_\textrm{dk}(\textbf{x},\textbf{y})$ is retained 
      provided that the condition
      \begin{equation}
      \label{eq:unweighting}
      \frac{d\sigma} {d\textbf{x}d\textbf{y}} \left( X_\textrm{dk}(\textbf{x},\textbf{y}) \right) > r W_{\textrm{max}}(X)
     \end{equation}
     is satisfied.  Otherwise
 the procedure to decay the event with kinematics $X$ is restarted 
      from step 1 onwards. This ensures that a decayed event 
      $X_{\textrm{dk}}(\textbf{x},\textbf{y})$
      is kept with a probability that is proportional 
to the differential cross section evaluated at this event.
\end{enumerate} 
It should be emphasized that only tree-level amplitudes are used 
when evaluating differential cross sections in the FLMW procedure. 
For production events with an extra hard radiation, 
spin correlation effects are included at next-to-leading-order accuracy.
For the other events, spin correlation effects are included at leading order 
accuracy, as no information on the one-loop corrections not factorizing the Born is used to generate 
the decay configurations.

Frixione \textit{et al.}~also recognized that, in the narrow width 
approximation, the maximum weight 
$W_{\textrm{max}}(X)$ can be expressed as a product
of the differential cross section 
for the undecayed process $d\sigma_{\textrm{no-dk}}/d\textbf{x}$ evaluated 
at $X$ and a constant factor $W^{\textrm{dk}}$ that is independent of 
the undecayed event $E_X$:
\begin{equation}
W_{\textrm{max}}(X) =
\frac{d\sigma_{\textrm{no-dk}}}{d\textbf{x}}(X)
\times W^{\textrm{dk}}
\label{eq:maxweight}
\end{equation}
They provided
analytical formulae for the constants $W^{\textrm{dk}}$'s associated with top quark and vector boson ($W^\pm,Z$) decays.
 
In the FLMW approach, off-shell effects are recovered 
by smearing the mass of each resonance according to a Breit-Wigner 
distribution. This requires to reshuffle the momenta of the 
external particles in the undecayed event. The procedure to perform
this momentum reshuffling is not specified in~\cite{Frixione:2007zp}.

The achievement in this paper is a generalization
of the procedure proposed by Frixione \textit{et al}.~to arbitrary 
processes, and the practical implementation in the \madgraph framework. 
This implementation implies that the code can be used 
for any processes 
for which the matrix elements can be evaluated 
with \madgraphb, so it applies to any model~\cite{Degrande:2011ua, deAquino:2011ub}.

On the conceptual level, generalizing the FLMW procedure
requires to solve two problems:
\begin{itemize}
\item a procedure to handle off-shell effects and momentum 
      reshuffling in a generic way must be established,

\item the maximum weight that is used for unweighting the events 
      [see Eq.~(\ref{eq:unweighting})] must be determined (at least numerically)
      on an automated basis.

\end{itemize}

We present our solutions for these problems in the two next 
sections, respectively.

\section{Off-shell effects and momentum reshuffling}

When scattering events with the production of heavy narrow resonances 
are simulated at next-to-leading-order accuracy, 
the width is typically set to zero, as it simplifies the 
calculation. 
However, the off-shellness of these resonances 
can have an non-negligible impact on the kinematics of the events.
As an example, in scattering events with the production of W bosons, 
the invariant mass of the decay products of the W boson lies 
outside the window $[m_W -10\textrm{ GeV}, m_W+
10 \textrm{ GeV}]$ with an $6.5 \%$ probability
(assuming a Breit Wigner distribution). Specific analyzes
may be sensitive to such effects, in which case off-shell effects 
must be incorporated, at least to some degree of accuracy.

Even though strictly speaking event generation can be factorized into a production
phase and a decay phase only in the narrow width approximation, prescriptions
can be used to recover part of the off-shell effects.
In the FLMW procedure, the mass of each resonance in undecayed events 
is smeared according to a Breit-Wigner distribution. 
The generated virtualities of the 
resonances enter explicitly in the set of variables 
$\textbf{y}$ characterizing the decay configuration, 
so that the unweighting procedure presented in Eq.~(\ref{eq:unweighting})
is used to capture any deviations with 
respect to a Breit-Wigner distribution
when decaying the events.
We adopt the same strategy in our procedure\footnote{It should be mentioned though that we do not 
consider effects from the partonic density functions when calculating the weight
associated with a decay configuration. This may lead to a small systematic bias in the 
distribution of  $2 \rightarrow 1$ s-channel events with respect to the invariant mass
of the s-channel resonance. 
}.

The off-shell effects imply that the masses of some 
external particles in the undecayed event with kinematics $X$ are altered, 
which requires to modify the external momenta 
 in such a way that  energy and momentum 
are still conserved. 
This modification of the momenta is sometimes called
momentum reshuffling. Practical implementations of momentum reshuffling rely on 
ad-hoc prescriptions. However, for a given process,  one strategy may be  better than
the others, as it could preserve some of the features of the distributions of events as predicted by the 
hard scattering amplitude associated with the production events.
This makes also clear the fact that the best approach to reshuffle momenta 
depends on the process under consideration.

In our algorithm, we pay particular attention to this aspect: momentum reshuffling is performed
in an optimized way by using diagram-based information of the tree-level scattering
amplitude associated with the undecayed events.
This is achieved by taking advantage of the single-diagram-enhanced multichannel
integration procedure in \madevent~\cite{Maltoni:2002qb}.
In that procedure, each channel of integration $i$
uses a phase-space mapping,
\begin{equation}
\label{eq:phasespacemapping}
\phi_i: \textbf{x}_i \rightarrow X(\textbf{x}_i ),
\end{equation}
that is linked to  a specific Feynman
diagram $i$. The connection between the diagram and the channel
is that each invariant associated with a propagator
 appearing in the diagram
is in one-to-one correspondence with a variable of integration in the channel.
 The amplitude  $A_i$ of the diagram associated with integration channel $i$
 is  used to set the relative weight $w_i$ of that channel point-by-point in the 
 phase space:
 \begin{equation}
\label{eq:weightchannel}
 w_i=\frac{|A_i|^2}{\sum_j |A_j|^2}.
 \end{equation}

 The connection with momentum reshuffling resides in the observation that a phase-space mapping
 $i$ as defined in Eq.~(\ref{eq:phasespacemapping})
 provides a convenient way to reshuffle the momenta.
The canonical variables $\textbf{x}_i=x_i^1, \dots , x_i^r$ characterizing the original event $X$
 are first determined
before smearing the masses of undecayed particles, by applying the 
inverse mapping $\phi_i^{-1}$ on the event with kinematics $X$. Then, after the masses are modified,
the kinematics of the reshuffled event $\tilde X$ is generated 
by applying the mapping\footnote{In some rare cases, this is kinematically impossible, 
we will comment on this problem later on.} 
 $\phi_i^*$ on the canonical numbers $\textbf{x}_i$
($\phi_i^{*}$ is the same mapping as $ \phi_i$ except for the masses of the 
undecayed particles, which are already determined in the set $\textbf{y}$). 

For each undecayed event  with the production kinematics $X$, 
our procedure to decay the event goes as follows.
\begin{enumerate}
\item For each channel of integration $i$ associated with the phase space of undecayed events,
   the squared amplitude of corresponding single diagram  $|A_i|^2$ is evaluated at the kinematics
   $X$ of the production event (before smearing the mass of the undecayed particles).                              
            
\item One channel of integration $i$ with mapping ($\phi_i: \textbf{x}_i \rightarrow X$) 
      is chosen randomly, 
with the probability to choose channel $i$ equal to $w_i$
given in Eq.~(\ref{eq:weightchannel}).
The inverse function $\phi_i^{-1}: X \rightarrow 
\textbf{x}_i$ is used to extract the canonical numbers $\textbf{x}_i$, which will remain unchanged when reshuffling the event.

\item The variables $\textbf{y}$ characterizing the decay are
  generated randomly, considering --for each decay-- a uniform
  distribution of the decay products in the rest frame of the
  decaying particle and a Breit-Wigner distribution for the
  virtuality of the resonance.

\item The momenta in the undecayed event
      are reshuffled by applying the phase-space 
      mapping $\phi_i^*(\textbf{y}): \textbf{x}_i \rightarrow \tilde{X}$ 
      on the canonical numbers $\textbf{x}_i$ extracted
      at step 2,
      where $\phi_i^*(\textbf{y})$ is the same mapping as 
      $\phi_i$ except that the pole mass of each final-state
      resonance has been replaced by the virtual mass generated
      at step 3 (hence its dependence on $\textbf{y}$).
      The resulting reshuffled event is denoted $\tilde{X}$.
       The momenta $\tilde{X}_\textrm{dk}(\textbf{x},\textbf{y})$ of the corresponding decayed event are reconstructed
       from the kinematics $\tilde{X}$, augmented with the subset of the variables $\textbf{y}$ that correspond to the angular distributions of the decay products in the rest frames of the resonances.

\item The differential cross section 
       $d\sigma / d\textbf{x}d\textbf{y}$
 associated with the decayed process is evaluated at $\tilde{X}_\textrm{dk}(\textbf{x},\textbf{y})$ 
      and an unweighting procedure with respect to 
      the maximum weight is used to decide whether this 
      decayed event should be kept or not.
      With $r$ a random number generated according to
       a flat distribution between 0 and 1, 
      the event is kept if
      \begin{equation}
      \label{eq:unweighting2}
      \frac{d\sigma} {d\textbf{x}d\textbf{y}}\left( \tilde{X}_{\textrm{dk}}(\textbf{x},\textbf{y}) \right) > r W_{\textrm{max}}(X)
     \end{equation}      
      where $W_{\textrm{max}}(X)$ is the maximum weight, of which the determination
      will be discussed in the next section.
      If the condition in (\ref{eq:unweighting2}) is not satisfied,
      the procedure to decay the event with kinematics $X$ is restarted 
      from step 3 onwards. 

\end{enumerate} 
One difficulty in this approach is that the mapping 
$\phi_i^*(\textbf{y}): \textbf{x}_i \rightarrow \tilde{X}$ that is used 
at step 4 is not always defined. This typically occurs
for events in which a pair of unstable particles has been generated
very close to threshold. If the difference between the invariant mass
$m_{12}$
of these particles and the sum of the pole masses $m_1+m_2$ is of the order 
of the width of one of the particles, 
the sum of the virtual masses may exceed the original invariant mass
in the event: $m_{12}<m_1^*+m_2^*$. If the invariant mass $m_{12}$
is mapped onto one of the canonical variables $\textbf{x}_i$,
the mapping 
$\phi_i^*(\textbf{y}): \textbf{x}_i \rightarrow \tilde{X}$ is ill-defined.
This indicates that 
the whole approach for factorizing the production of undecayed events
from the decay cannot be used to get the fine details of 
event distribution very close to threshold. This is no surprise, because in this region of phase space the narrow width approximation that was used to generate the undecayed events breaks down.
In our implementation, if the mapping 
$\phi_i^*(\textbf{y}): \textbf{x}_i \rightarrow \tilde{X}$ is not defined, 
the procedure is restarted from step 3 onwards. This may lead to
a migration of events very close to a threshold to events just below 
the threshold. This effect is negligible in the limit $\Gamma_i/m_i \rightarrow 0$.

One main advantage of the procedure outlined above is that if resonance effects 
are present in the production events, those effects are systematically
preserved in spite of the momentum reshuffling  (due to step 2). Indeed, an invariant mass
which is resonant 
is systematically mapped onto a variable of integration in all the dominant channels
for events in the resonance region. Then, by construction, this invariant mass is not affected 
by the reshuffling procedure.

\section{Estimation of the maximum weight}

The unweighting procedure outlined in Section~\ref{sec:spin} 
requires an upper bound on the differential cross section, 
given by the maximum weight $W_{\textrm{max}}(X)$.
In the work of~\cite{Frixione:2007zp}, 
the maximum weight was determined analytically
for decays of weak bosons and the top quark by
calculating the constant $W^{\textrm{dk}}$ in Eq.~(\ref{eq:maxweight})
for each of these decay chains.
In this work we opted for a numerical estimate
of the quantity $W^{\textrm{dk}}$. 
Such a numerical estimate of the maximum weight is
 better suited to automate 
the FLMW procedure to any given process, in particular for New Physics Models.

The simplest solution to get a numerical 
estimate of the maximum weight is to probe the phase space of the decay 
of the first production event with a large number of points $N$. 
The largest value of the differential 
cross section for the decayed process normalized by the 
differential cross section for the production process
provides a quantity $W_{1}^{\textrm{dk}}$
(the index $1$ refers to the \emph{first} production event)
that approximates the required upper bound on the maximum weight.
However, such an estimate fails to account
for finite width
corrections to Eq.~(\ref{eq:maxweight}), 
and it also
suffers from large statistical
uncertainties (unless the number of phase-space points $N$
 is really large).
Our prescription 
is slightly more elaborate to deal with these two issues:
\begin{itemize}
\item we extract the estimates   
$W_{1}^{\textrm{dk}}, \dots, W_{m}^{\textrm{dk}} $
for the first $m$ production events, 

\item we express the estimate of an upper bound on $W^{\textrm{dk}}$, denoted by $W_{\textrm{max}}^{\textrm{dk}}$, 
in terms of the mean value $\left< W_{i}^{\textrm{dk}} \right>$
and the standard deviation $\mathrm{std}\left( W_{i}^{\textrm{dk}} \right)$ 
associated with the numbers $W_{1}^{\textrm{dk}}, \dots, W_{m}^{\textrm{dk}} $:
\begin{equation}
  \label{eq:numerical-algorithm-Wmax}
  W_{\textrm{max}}^{\textrm{dk}} = \left< W_{i}^{\textrm{dk}} \right> + \xi \,\, \mathrm{std}\left( W_{i}^{\textrm{dk}} \right) .
\end{equation} 
\end{itemize}
In this way, we account for finite width corrections by considering multiple events, 
since those corrections depend on the kinematics of the production event.
Statistical uncertainties as well as fluctuations due to finite width corrections are minimized by using 
the average and standard deviation in Eq.~(\ref{eq:numerical-algorithm-Wmax}), 
rather than (for example) the largest of all $W_i^{\textrm{dk}}$.
This results in a reliable unweighting efficiency.

Our prescription contains three adjustable parameters:
the number of phase-space points $N$ associated with the decay,  
the number of production events $m$, and the prefactor $\xi$
in Eq.~(\ref{eq:numerical-algorithm-Wmax}).
These parameters were calibrated so that the quantity 
 $W_{\textrm{max}}$ is as close
as possible to (while strictly larger than) the true value of the maximum weight,
in order to optimize the Monte Carlo unweighting efficiency.
Based on empirical studies involving different processes,
we verified that the same calibration can be used for all processes.

\begin{figure}
\includegraphics[scale=0.6]{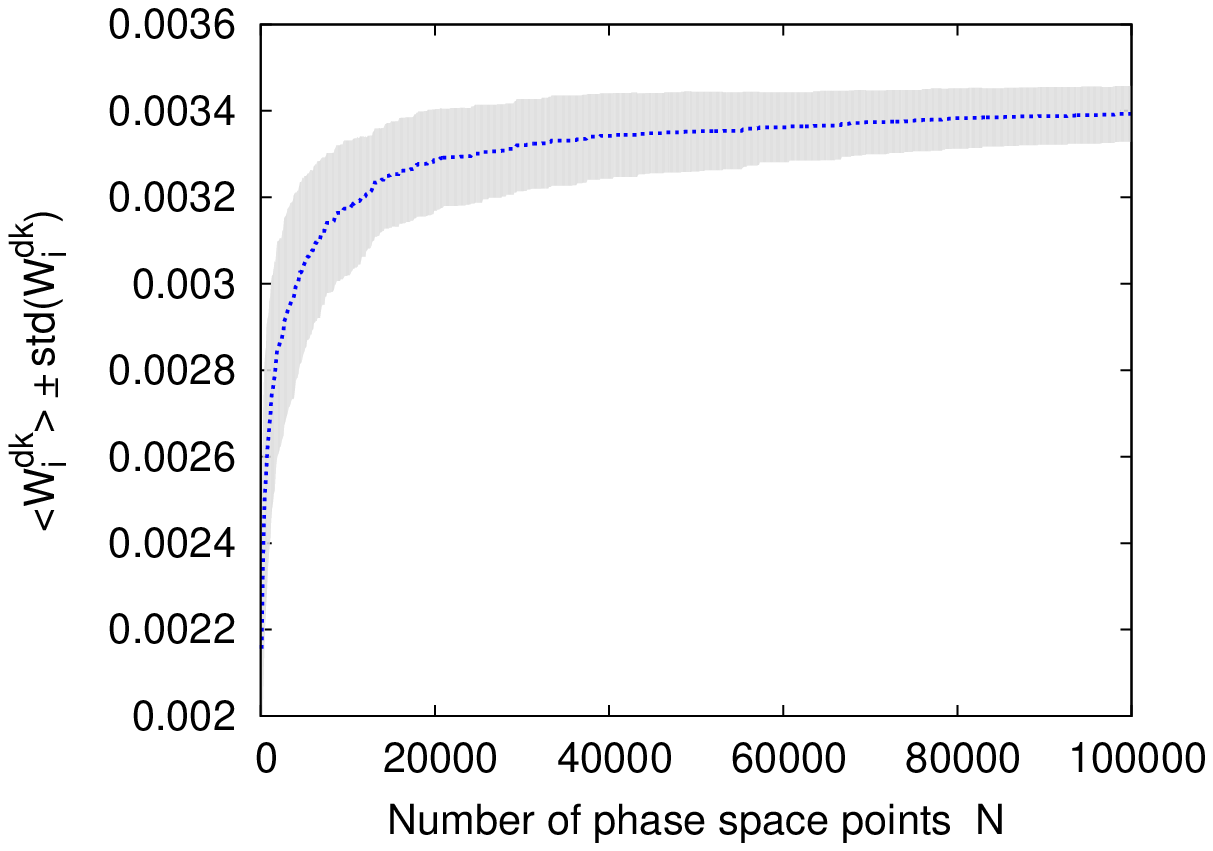}
\includegraphics[scale=0.6]{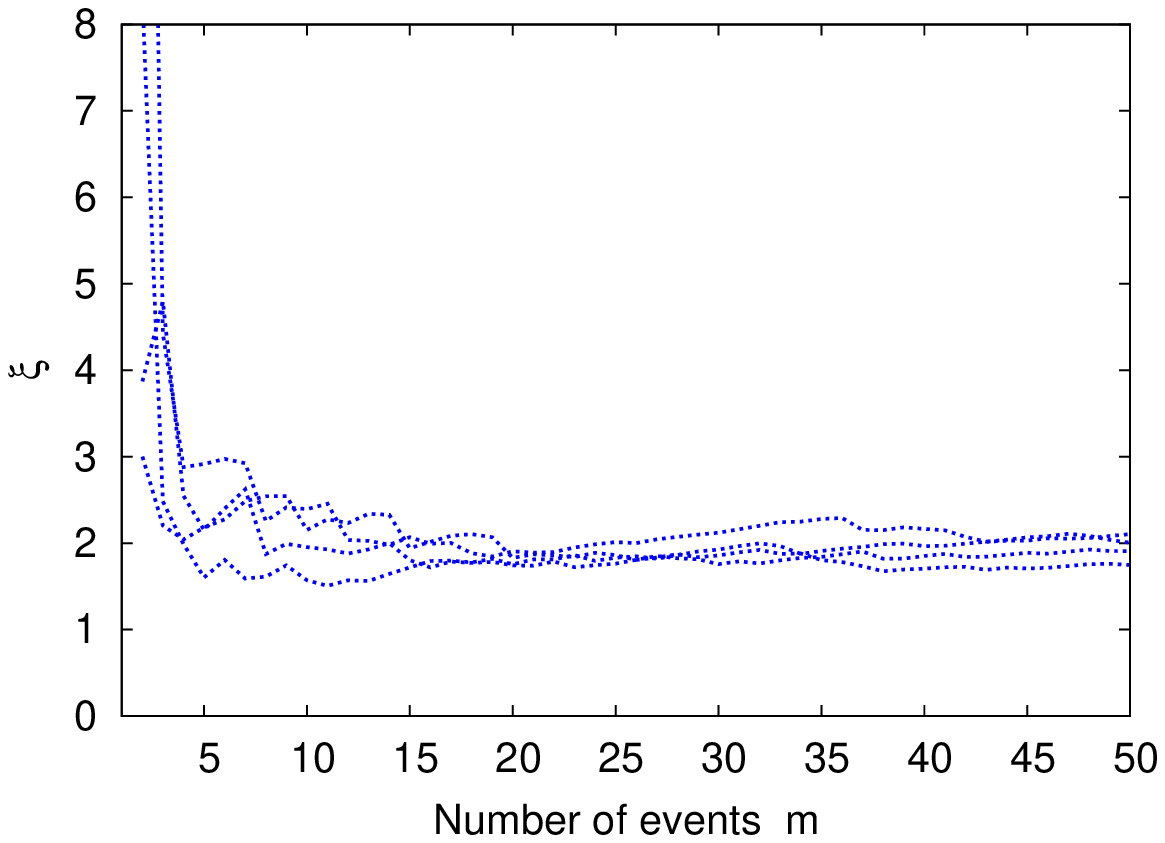}
\caption{
Calibration of the relevant parameters in the numerical estimate of the maximum weight, based on $pp \rightarrow W$ events.
Left pane: the average and standard deviation of the $W_i^{\textrm{dk}}$ estimate from $100$ production events 
are shown as
a function of the number of phase-space points $N$. Right pane: the parameter $\xi$ extracted to reproduce the
true value of the maximum weight is shown as a function of the number of production events $m$ (the number
of phase-space points $N$ is fixed to $20$k). 
The different curves are associated with different sets of production events.
\label{fig:maxweight}
}
\end{figure}

In Figure~\ref{fig:maxweight}, we illustrate the calibration 
of the three parameters $N,m$ and $\xi$ based on $p p \rightarrow W$ events.
First, the number of phase-space points $N$ is increased 
until the average of the $W_i^{\textrm{dk}}$ is constant 
with respect to the number of phase-space points\footnote{
While the average becomes constant at large $N$, the $W_i^{\textrm{dk}}$ 
do fluctuate event-by-event due to finite width corrections.
This is displayed by the nonzero standard deviation at large $N$.} (left pane).
This identifies the number of phase-space points to be used.
Next, the true value of $W^{\textrm{dk}}$ is extracted 
by considering a very large number of production events.
Once this value is known, the prefactor $\xi$ in 
Eq.~(\ref{eq:numerical-algorithm-Wmax}) can be extracted 
for a given number $m$ of production events so that 
the resulting estimate  $W_{\textrm{max}}^{\textrm{dk}}$ reproduces
the true value $W^{\textrm{dk}}$ (right pane). 
We take the number of production events $m$ large enough 
so that the associated value of $\xi$ converges to a constant value. 
That constant then gives the value of $\xi$, 
thereby fixing the remaining parameter in our prescription.
In the code, we have made the conservative choice 
$(N=10^4,\,m=20,\,\xi=4)$. 

\section{Validation}

In order to support phenomenological studies in an optimal way, 
Monte Carlo generators must show a high level of efficiency, 
while reflecting as much as possible the accuracy of state-of-the-art calculations.
Regarding the efficiency, the approach discussed in this paper
has been implemented in the \madgraph  framework, and hence
it can be used for a large class of processes within or beyond
the standard model in a complete automated way. 
Moreover, since the decay of a specific event typically requires 
only a few evaluations of tree-level matrix elements, 
generating the decay is in general fast. 
The efficiency and the flexibility of the tool
is at the cost of two approximations inherent to
the procedure at work: 
\begin{enumerate}
\item some finite-width effects
may be lost in the distributions of events, as the 
procedure is built upon the narrow width approximation
in the first place, 
\item only tree-level matrix-elements
are used to calculate the weight associated with 
a specific decay configuration.
\end{enumerate}
We comment on the validity 
of these two approximations in this section.

\subsection{Finite width effects}

\begin{figure}
\includegraphics[scale=0.6]{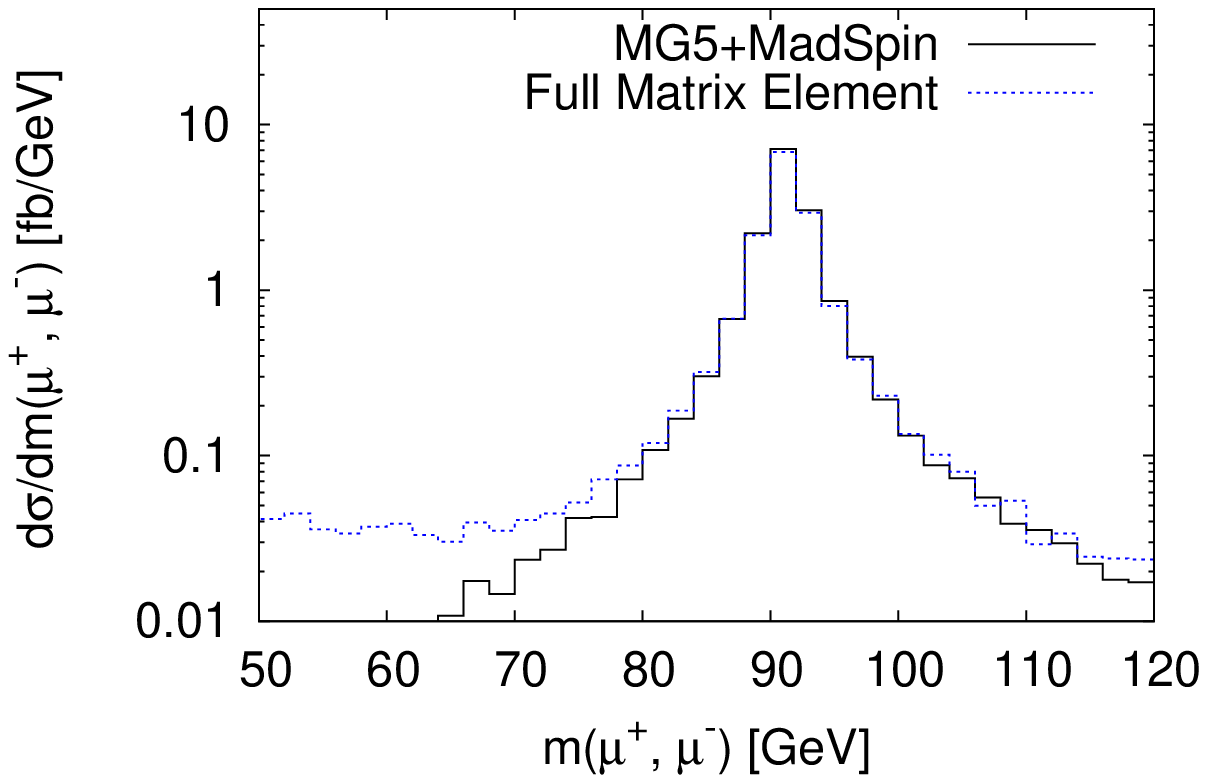}
\includegraphics[scale=0.6]{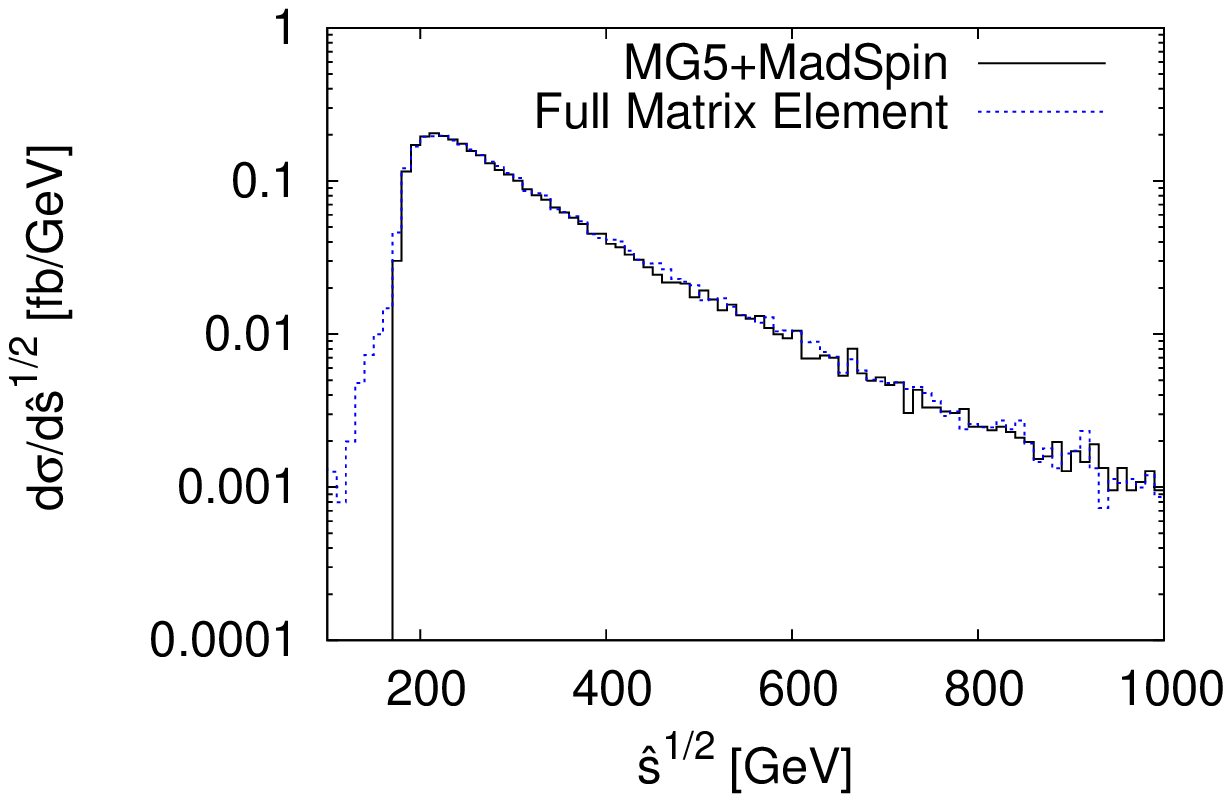}
\caption{
Distribution of events with respect to the invariant mass of the muon pair (left pane) and 
with respect to $\sqrt{\hat{s}}$ (right pane) in   
$pp \rightarrow  \mu^+ \mu^- e^+ \nu_e$ events.
The two histograms result from two distinct procedures to generate the events:
either \madgraph is used 
to generate $pp \rightarrow  Z W^+ $ (LO accuracy) which 
are subsequently decayed using \madspin
(solid histogram) or  \madgraph is used to the generate $pp \rightarrow 
\mu^+ \mu^- e^+ \nu_e $ in one shot, so that 
all finite-width effects are systematically included (dashed histogram).
\label{fig:ZW}
}
\end{figure}

We verified the validity of the first approximation
by considering a large class of tree-level processes 
involving heavy resonances. We generated events in two different 
ways (a) we used \madgraph to generate undecayed events, which were
subsequently decayed using \madspinb,  
(b) we used \madgraph to generate decayed events, considering
a finite width for the heavy resonances at all stages of the generation
(and including also the non-resonant contributions).
We compared a large number 
of distributions involving the transverse momentum, the angular separation,
and the invariant mass of the final state particles. We obtained  very good 
agreement between distributions resulting from the generation procedures (a) and (b). 
Some differences were observed in the tails 
of resonant invariant mass distributions or 
close to threshold, as expected.

To illustrate the largest observed deviations, let us consider the specific example of
diboson production at the LHC in the channel 
$pp \rightarrow ZW^+ \rightarrow \mu^+ \mu^- e^+ \nu_e$.
Figure~\ref{fig:ZW} shows the distribution
of events with respect to the invariant mass of the muon pair (left pane) and 
with respect to $\sqrt{\hat s}$ = the invariant mass of the colliding partons
 (right pane) resulting 
from the previously-mentioned 
procedures  to generate  events: (a) \madgraph is used 
to generate $pp \rightarrow Z W^+$ events (at leading order accuracy) which 
are subsequently decayed using \madspin (solid histogram), 
(b) \madgraph is used to the generate $pp \rightarrow 
\mu^+ \mu^- e^+ \nu_e $ events in one shot
 (including also the non-resonant diagrams with the photon splitting $\gamma^* \rightarrow \mu^+ \mu^-$), 
so that 
all finite-width effects are systematically included (dashed histogram). 
In this last case, we impose the cut $m(\mu^+, \mu^-) > 40$ GeV. 

As expected,  procedure (a) fails to reproduce the correct
 distribution of events  
far away from the resonance region $m(\mu^+, \mu^-)\approx m_Z$, 
as  the distribution of events in these
regions is sensitive to the non-resonant contributions involving the 
photon splitting $\gamma^* \rightarrow \mu^+ \mu^-$.
We observe though that the distribution of events 
with respect to the invariant mass of the muon pair 
is accurately reproduced in a rather extended  region
around the pole mass $m_Z$:  
 although the Z boson  is generated 
on its mass shell in undecayed events, off-shell effects are recovered
to a very good accuracy when decaying the events.
We also observe a good agreement for the distribution of  events 
with respect to the invariant mass of the 
colliding partons ($\sqrt{\hat s}$), except
below the threshold region $\sqrt{\hat s}\approx m_Z+m_W$
where the effects from the finite widths of the Z and W
bosons are of primary importance and cannot be reproduced 
in the narrow width approximation.
An ad-hoc approach to improve the description below the $m_Z+m_W$
threshold would be to allow for a (small) change in $\sqrt{\hat s}$
when the ratio $(m_Z+m_W)/\sqrt{\hat s}$ is close to one.
We leave this for future work.

\subsection{Spin correlation effects in NLO events}

\label{sec:spin}

\begin{figure}[t!]
\includegraphics[scale=0.6]{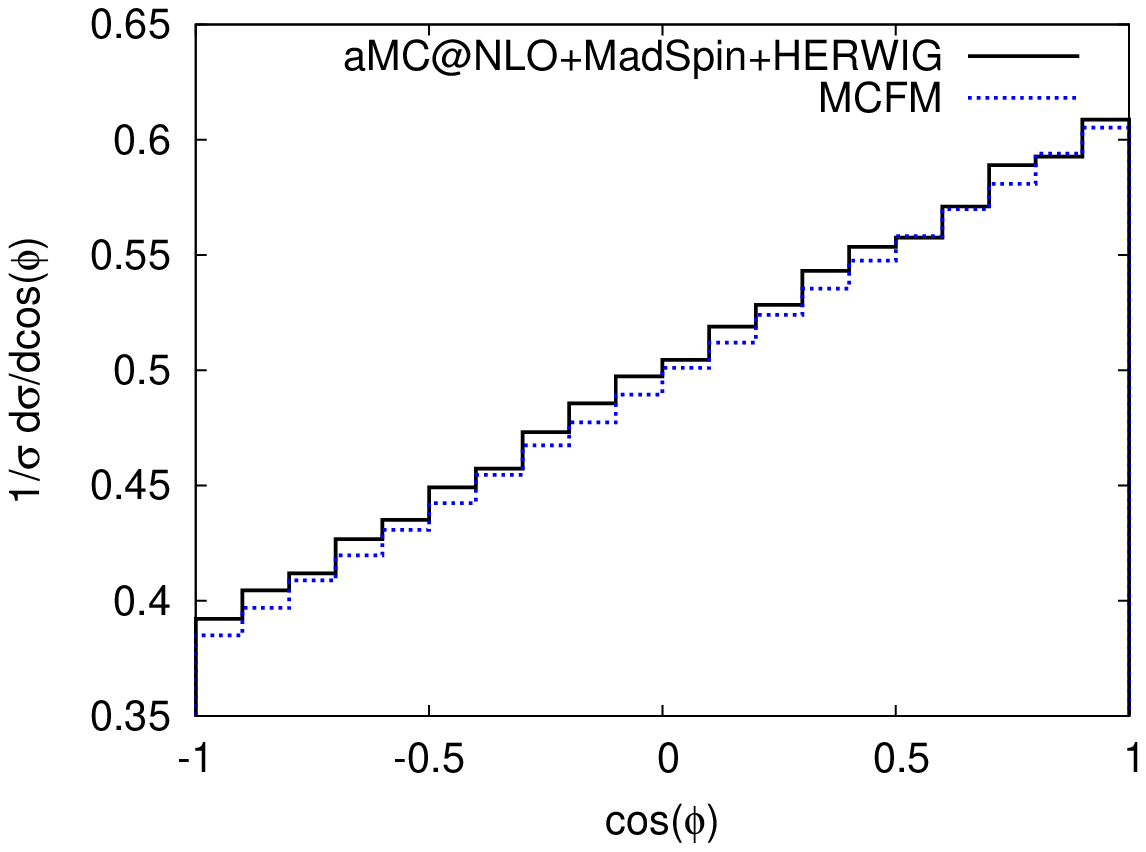}
\includegraphics[scale=0.6]{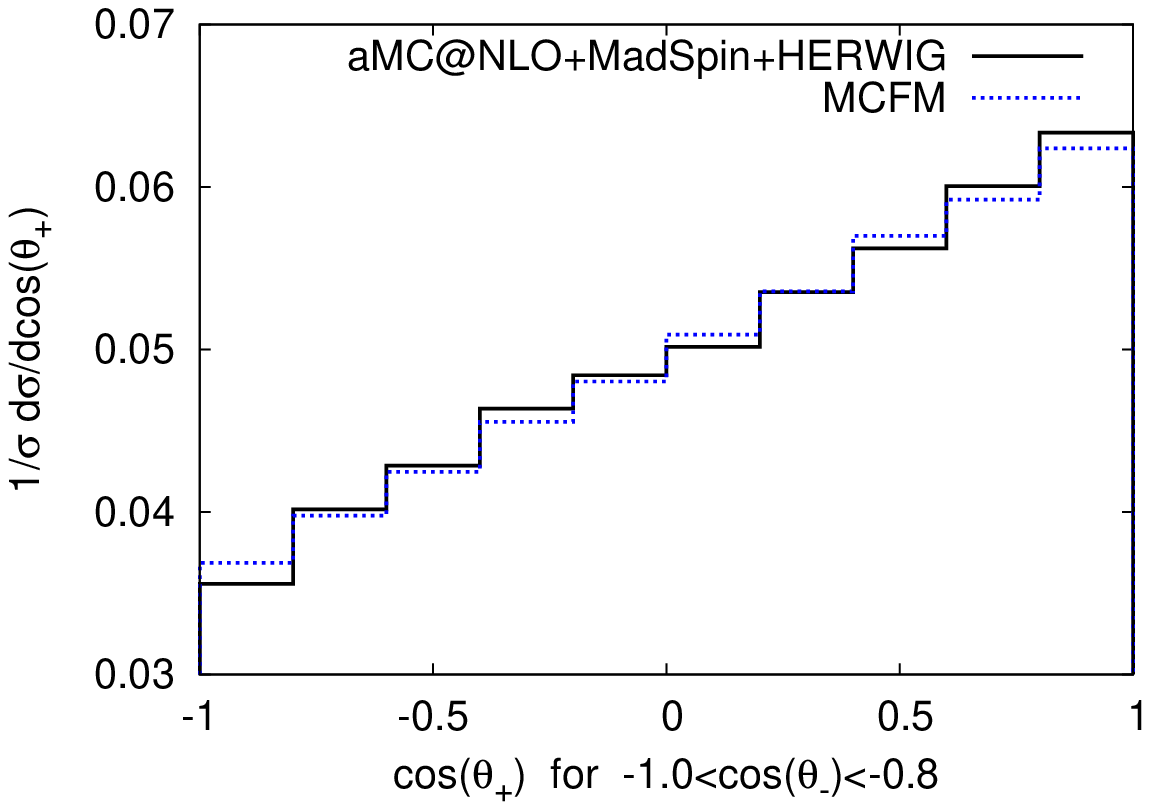}
\includegraphics[scale=0.6]{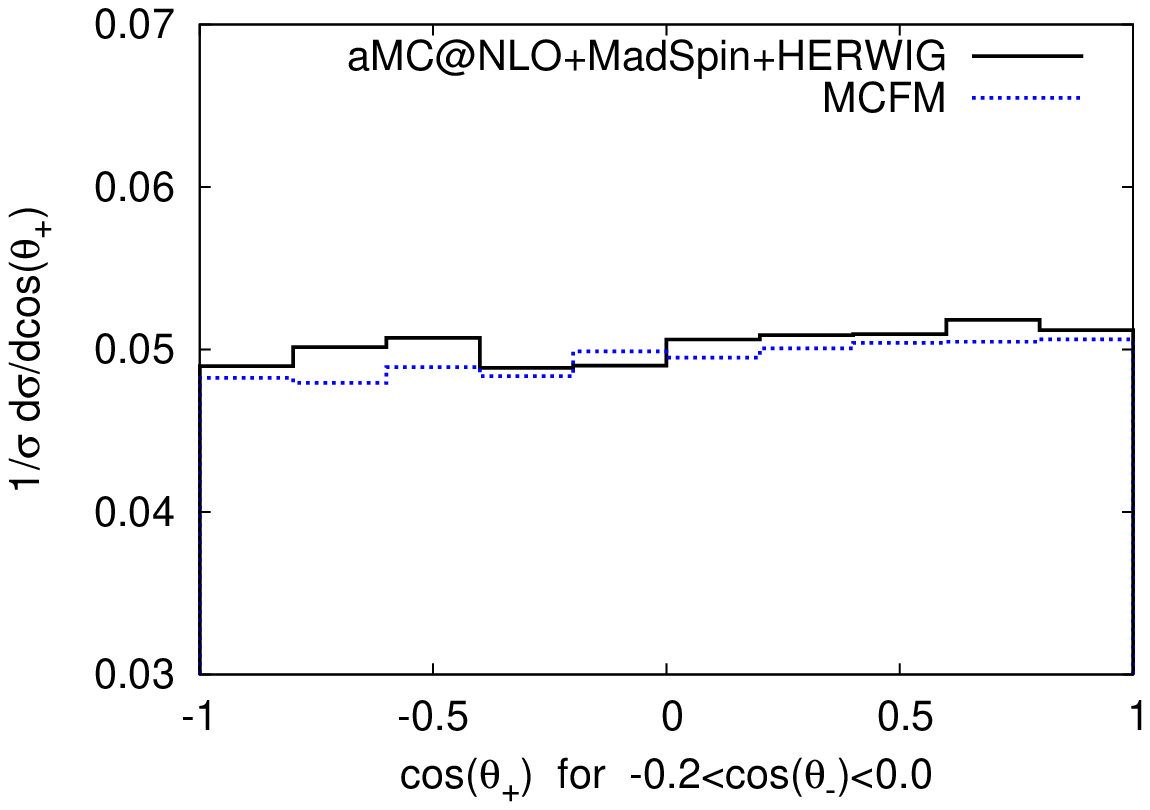}
\includegraphics[scale=0.6]{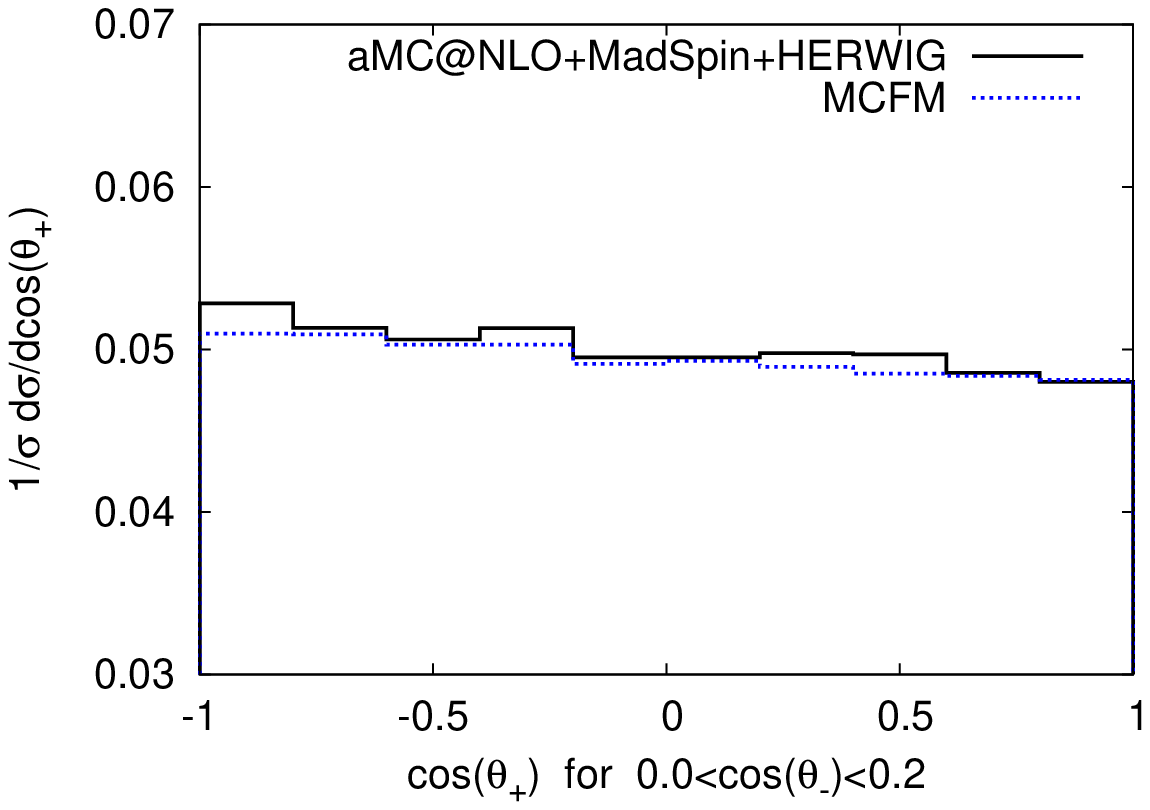}
\includegraphics[scale=0.6]{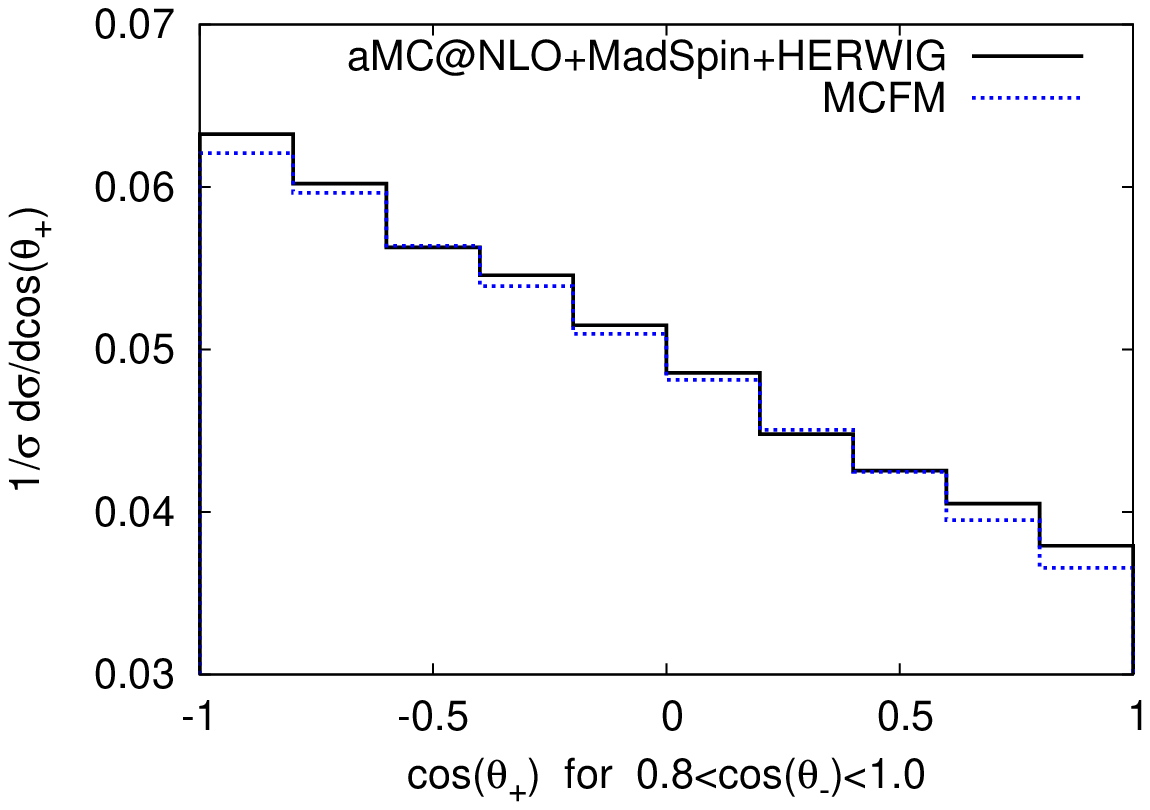}
\includegraphics[scale=0.6]{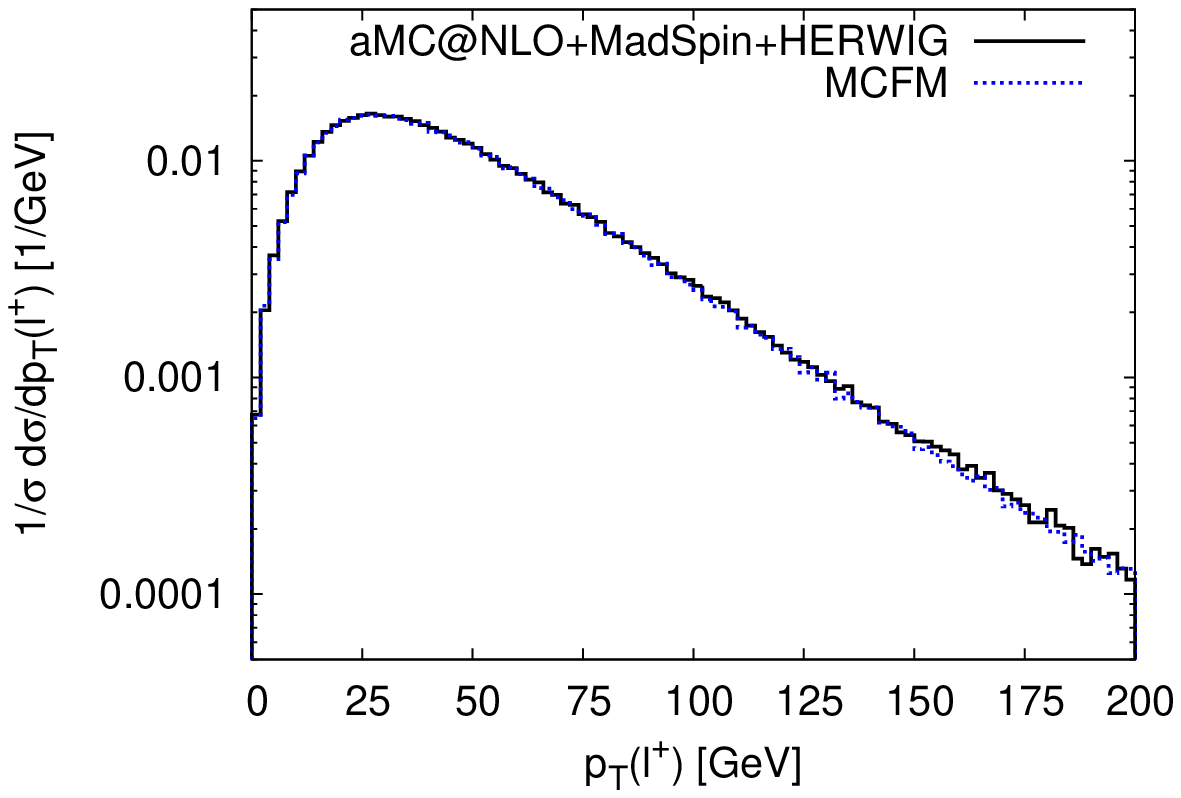}
\caption{Next-to-leading-order cross sections
differential in $\cos \phi$, $\cos (\theta_+)$
and $p_T(l^+)$ for  $p p \rightarrow t \bar t $
events in the dileptonic channel. The   angles are defined
in the text.
}
\label{fig:tt_distr}
\end{figure}

For the decay of Les Houches Events generated at next-to-leading-order accuracy, 
the procedure in \madspin retains spin correlation effects at tree-level accuracy, 
\ie no information from the virtual amplitude is used in calculating the weight
of a specific decay configuration. 
The validity of this second approximation
can be assessed in some specific cases by comparing distributions 
of events against predictions
including spin correlation effects at a higher level of accuracy.
One possibility is to generate these predictions with \MCFM 
(provided that it includes the process at work)
which 
includes all QCD corrections 
when calculating the weight of a decay configuration.
A limitation of such a comparison is that \MCFM provides only
fixed-order predictions, whereas \madspin is embedded in a scheme
involving NLO calculation matched to parton shower.
A more advanced validation procedure 
(\eg involving NLO predictions in the complex mass scheme matched 
to parton shower) is beyond the scope of this paper.

In this section, we present  comparison plots with the results from  the Monte Carlo program
 \MCFM
for $t \bar t$ and single-$t$ production at the LHC at $\sqrt{s} = 8 \mathrm{\,TeV}$.
For single-$t$~\cite{Campbell:2009ss} we used the four-flavor scheme and considered production via the $t$-channel.
In both processes the $t$ and $\bar t$ decay semi-leptonically (b quark + lepton + neutrino). 
 QCD corrections in the decay itself are ignored.
The relevant masses are $m_t = 172.5 \GeV$ and $m_b = 4.75 \GeV$.
Jets are reconstructed by means of the anti-$k_T$ algorithm~\cite{Cacciari:2008gp}, with $R_{\mathrm{cut}} = 0.4$ and $(p_{T})_{\mathrm{min}} = 25 \GeV$.
For $t \bar t$ we have used the first PDF set from MSTW2008nlo68cl (v5.7)~\cite{Martin:2009iq}.
In that case $\alpha_S(M_Z) = 0.12018 $, which evolves by two-loop running.
For single-$t$ we used MSTW2008nlo68cl\_nf4 (v5.8.4) instead~\cite{Martin:2010db}, where $\alpha_S(M_Z) = 0.11490 $.
The renormalization and factorization scales are set equal to the same scale $\mu$.
In the case of $t \bar t$, $\mu$ is given by the average of the $t$ and $\bar t$ transverse masses, 
while in the case of single-$t$ we use four times the $b$ transverse mass, 
this choice being based on the work of~\cite{Frederix:2012dh}.
In both cases, top quarks are decayed in the semi-leptonic channel.
Events are showered and hadronized with the program \herwig\cite{Corcella:2000bw}. 

\begin{figure}[t!]
\includegraphics[scale=0.6]{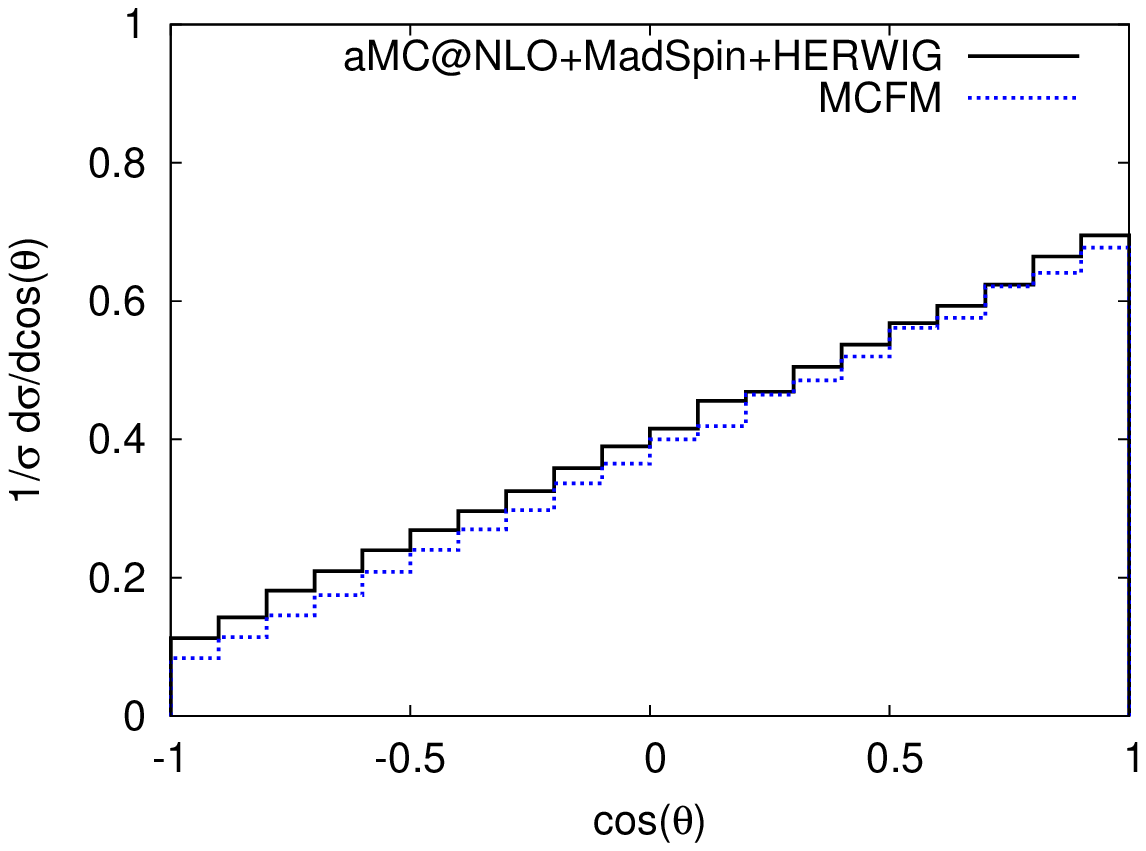}
\includegraphics[scale=0.6]{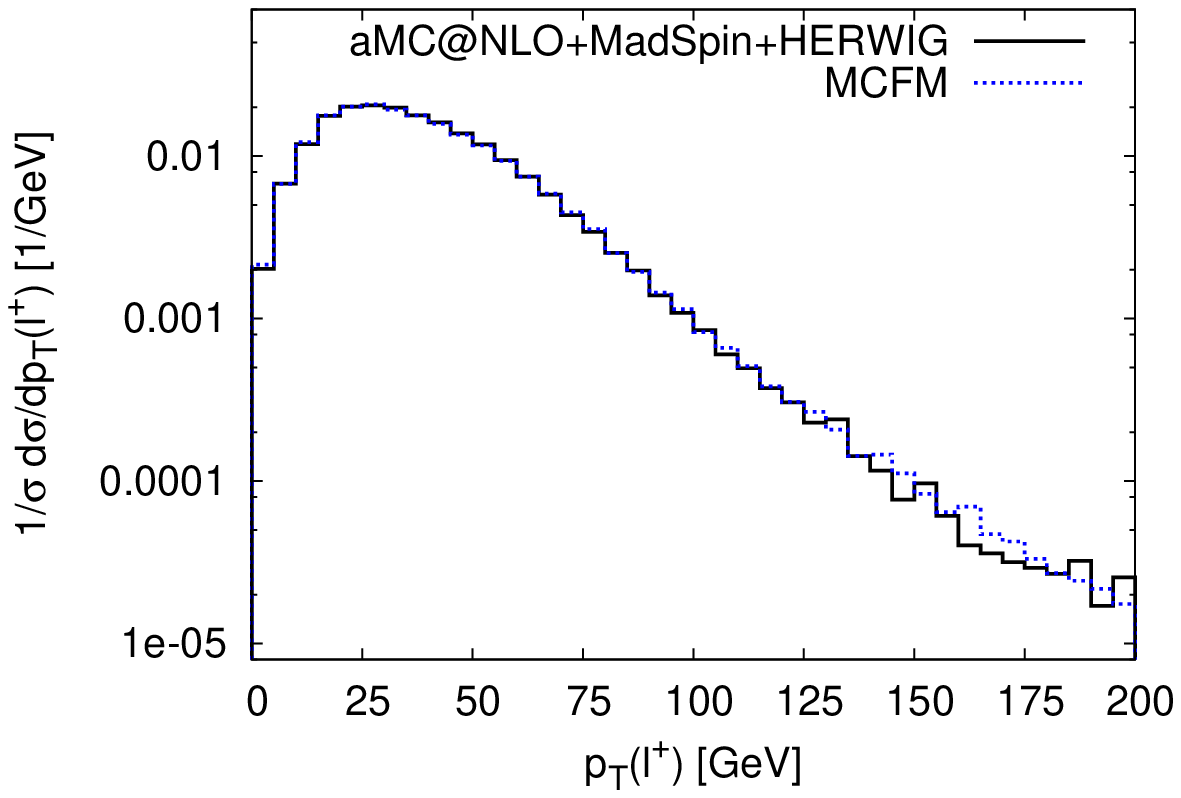}
\caption{
Next-to-leading-order cross sections 
differential in $\cos (\theta)$ and
 $p_T(l^+)$ for  single-top production in the four 
flavor scheme (t-channel only).  The 	angle $\theta$ is defined 
in the text.  
}
\label{fig:singlet_distr}
\end{figure}

For top quark pair production, the observables that are most sensitive 
to spin correlation effects were discussed in~\cite{Bernreuther:2004jv}:
\begin{itemize}
 \item $\cos(\phi)$, where $\phi$ is the angle between the direction of flight of $l^+$ in the $t$ rest frame and the direction of flight of $l^-$ in the $\bar t$ rest frame,
 \item $\cos(\theta_\pm)$, where $\theta_+$ ($\theta_-$) is the angle between the direction of flight of $l^+$ ($l^-$) in the $t$ ($\bar t$) rest frame and positive beam direction.
\end{itemize}
As can be seen in Figure~\ref{fig:tt_distr}, predictions for these observables 
using the scheme outlined in this
paper (\amcatnlob+\madspinb+\herwigb) are overall in very good agreement with those generated by \MCFMb. 
A good agreement is also found for the $p_T$ spectrum of the positively-charged lepton 
[Figure~\ref{fig:tt_distr}, bottom right].

For single-top production, the observables that are most sensitive 
to spin correlation effects were discussed in~\cite{Mahlon:1996pn}.
In the case of t-channel production, the angle $\theta$
is defined in the $t$ rest frame as the angle between the directions 
of flight of $l^+$ and the hardest non-$b$-tagged jet.
In this case as well,  predictions using the scheme outlined in this
paper (\amcatnlob+\madspinb+\herwigb) are  in good agreement with those generated by \MCFM
(see Figure~\ref{fig:singlet_distr}). 
The apparent difference in normalization of the two curves in the plot on the left-hand side of
Figure~\ref{fig:singlet_distr} is due to the definition of the observable. For the \amcatnlob+\madspinb+\herwig predictions there are slightly
more events with a non-$b$-tagged jet compared to the fixed order \MCFM results (80\% versus 76\%, respectively).


\section{Application: top-quark pair production in association with a light Higgs boson }

In order to illustrate the capabilities of the tool, we apply it to
the case of top-quark pair production in association with a light
Higgs boson at the LHC (running at 8 TeV), considering  both the 
scalar and pseudo-scalar hypotheses for the Higgs boson.
Due to the large irreducible QCD background, 
any search strategy for this Higgs production process
relies strongly on the accuracy of the Monte Carlo predictions.
QCD correction to these
processes has been analyzed by two
groups~\cite{Frederix:2011zi,Garzelli:2011vp} and a comparison between
these independent calculations has appeared in
Ref.~\cite{Dittmaier:2012vm}. In these works it was shown that the NLO
corrections are very mild, in particular on shapes of distributions.
 
To the best of our knowledge, the problem of retaining spin
correlation effects in events generated at NLO accuracy has not been
addressed yet for these processes. This problem is trivially solved
using the scheme proposed in this paper: NLO parton-level events are
generated with \amcatnlob, (LHC at 8 TeV, $\textrm{PDF set} =
\textrm{MSTW2008(n)lo68cl}$, $m_H=m_A=125$ GeV,
$\mu_R=\mu_F=(m_T(H/A)m_T(t)m_T(\bar{t}))^{(1/3)}$, no cuts) and then
decayed with \madspin before they are passed to \herwig for shower and
hadronization.  In this illustration, top and anti-top quarks are
decayed semi-leptonically, whereas the Higgs is decayed into a pair of
b quarks.

\begin{figure}[t!]
\includegraphics[scale=0.6]{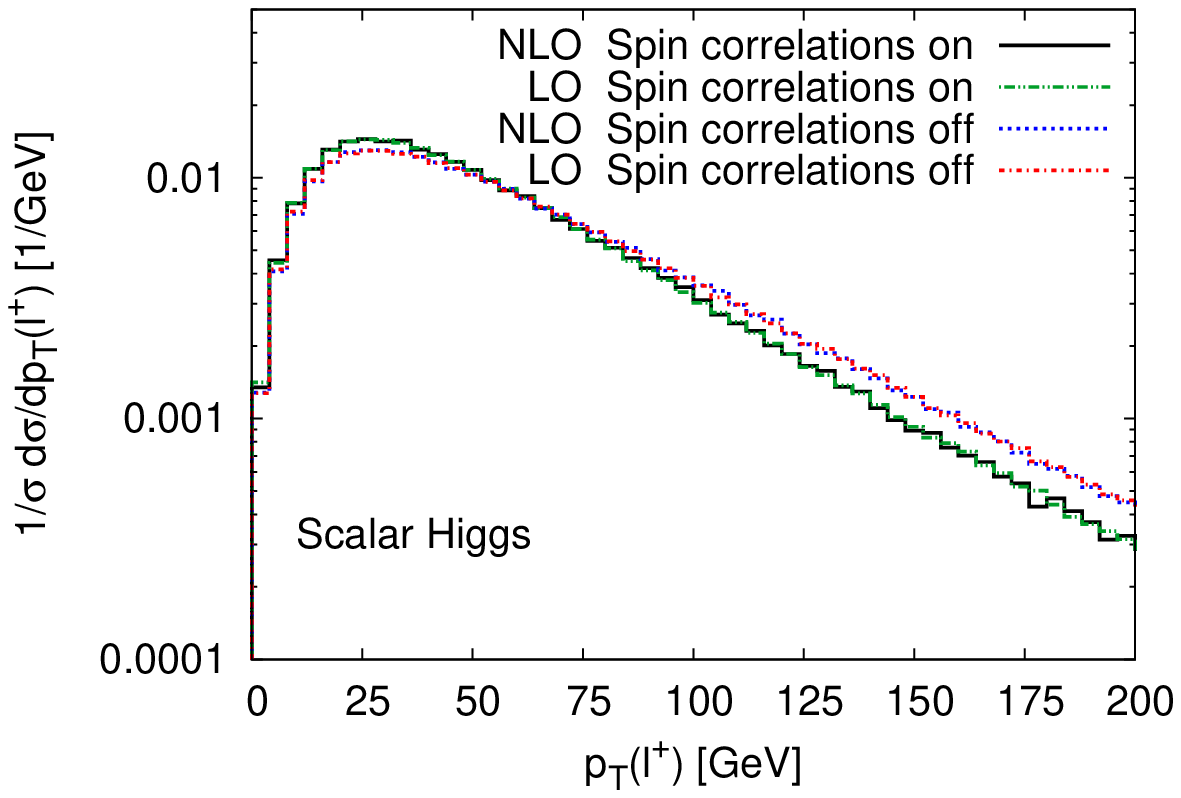}
\includegraphics[scale=0.6]{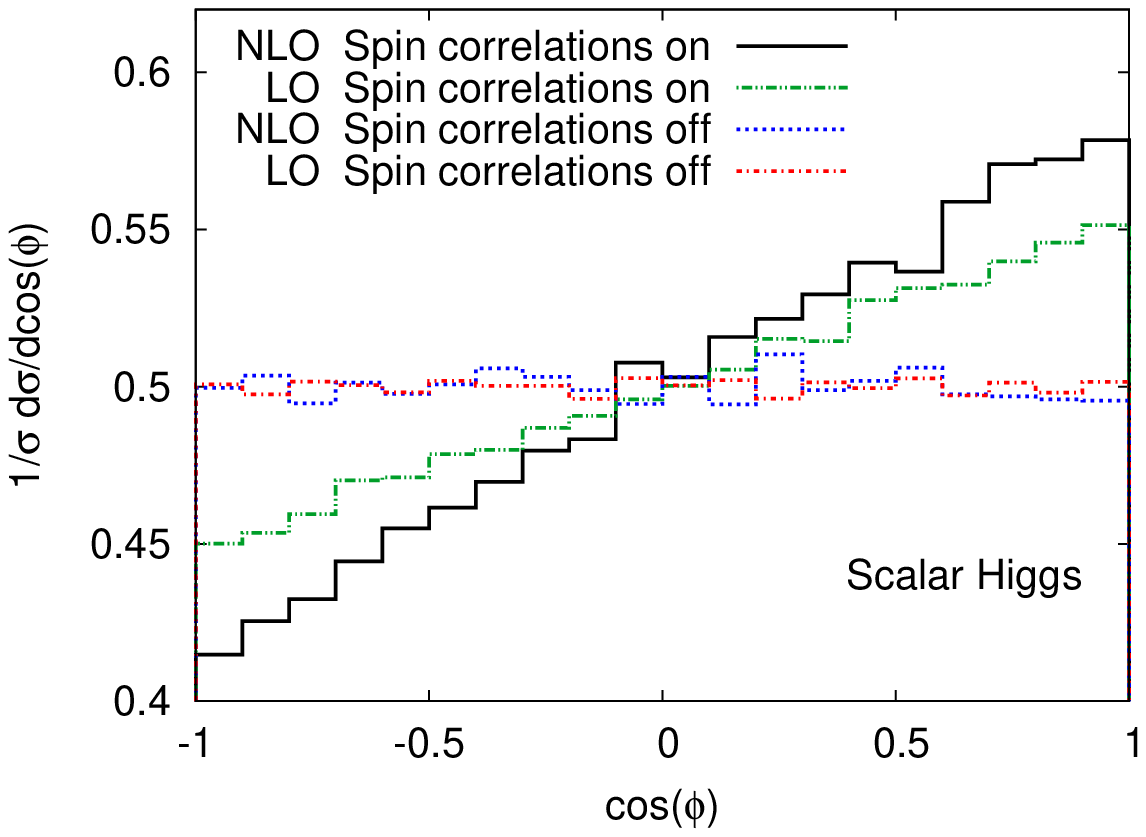}
\caption{Next-to-leading-order cross sections differential in
  $p_T(l^+)$ (left pane) and in  $\cos \phi$ (right pane) for $t
  \bar t H$ events with or without spin correlation effects. For
  comparison, also the leading-order results
  are shown.  Events were generated with \amcatnlob, then
  decayed with \madspinb, and finally passed to \herwig for shower and
  hadronization.  }
\label{fig:htt_distr}
\end{figure}

Figure~\ref{fig:htt_distr} shows the normalized distribution of events
with respect to $\cos (\phi)$ (which was defined in the previous
section), and with respect to the transverse momentum of the hardest
positively-charged lepton. Although spin correlation effects
significantly distort the distribution of events with respect to $\cos
(\phi)$, their impact on the $p_T$ spectrum of the leptons is milder,
except at large transverse momentum. The relatively larger effect in
the tail of this distribution can easily be understood from the fact
that the inclusion of the spin correlations is a unitary procedure: a
small change at low $p_T$, where the cross section is large, needs to
be compensated by a larger (relative) effect at high $p_T$.

It is interesting that spin correlations have a much more dramatic
influence on the shape of the $p_T$ spectrum than  NLO
corrections: the leading order results fall directly on top of the NLO
results for these normalized distributions (both without spin
correlations), as it can be seen by comparing the dotted blue and
dash-dotted red curves. This suggests that preserving spin
correlations is more important than including NLO corrections for this
observable. However, we observe that the inclusion of both,
as it is done here, is necessary for an accurate prediction of the 
distribution of events with respect to $\cos
(\phi)$. In general, a scheme including both spin correlation 
effects and QCD corrections is preferred: 
it retains the good features of a NLO calculation,
\ie reduced uncertainties due to scale dependence (not shown), while
keeping the correlations between the top decay products.

\begin{figure}[t!]
\includegraphics[scale=0.6]{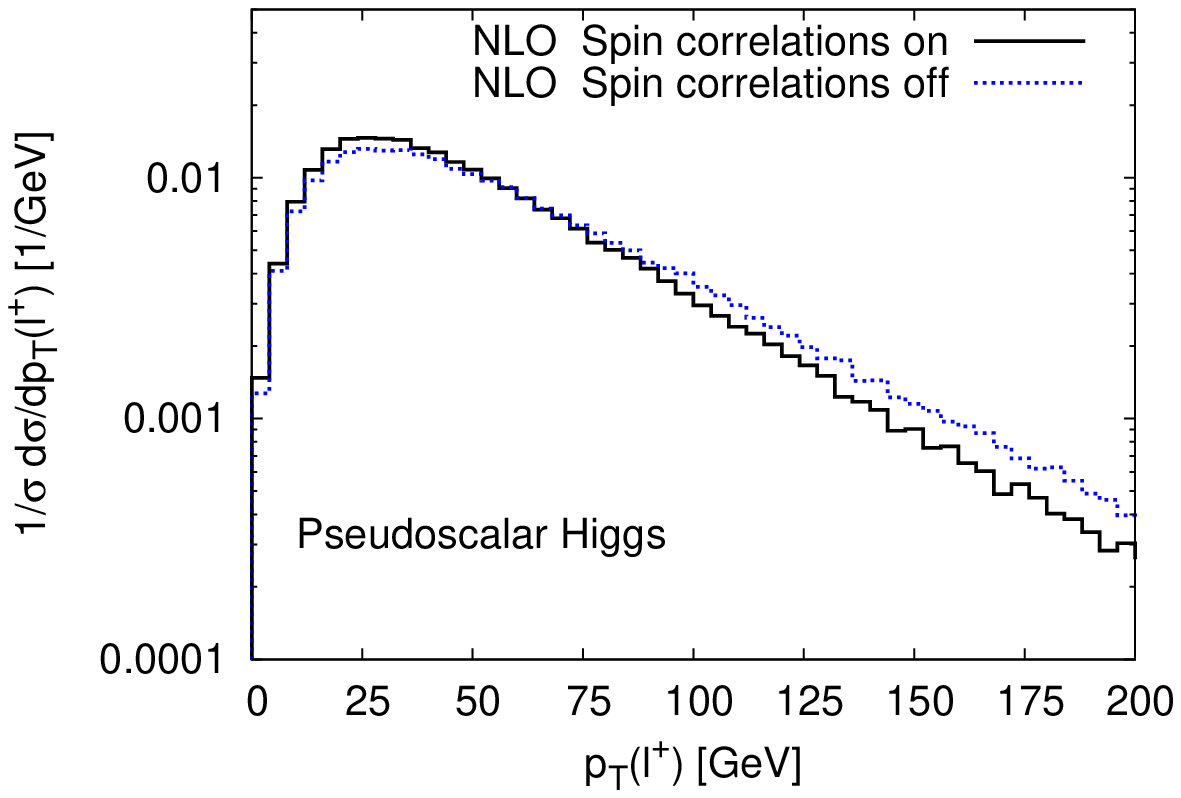}
\includegraphics[scale=0.6]{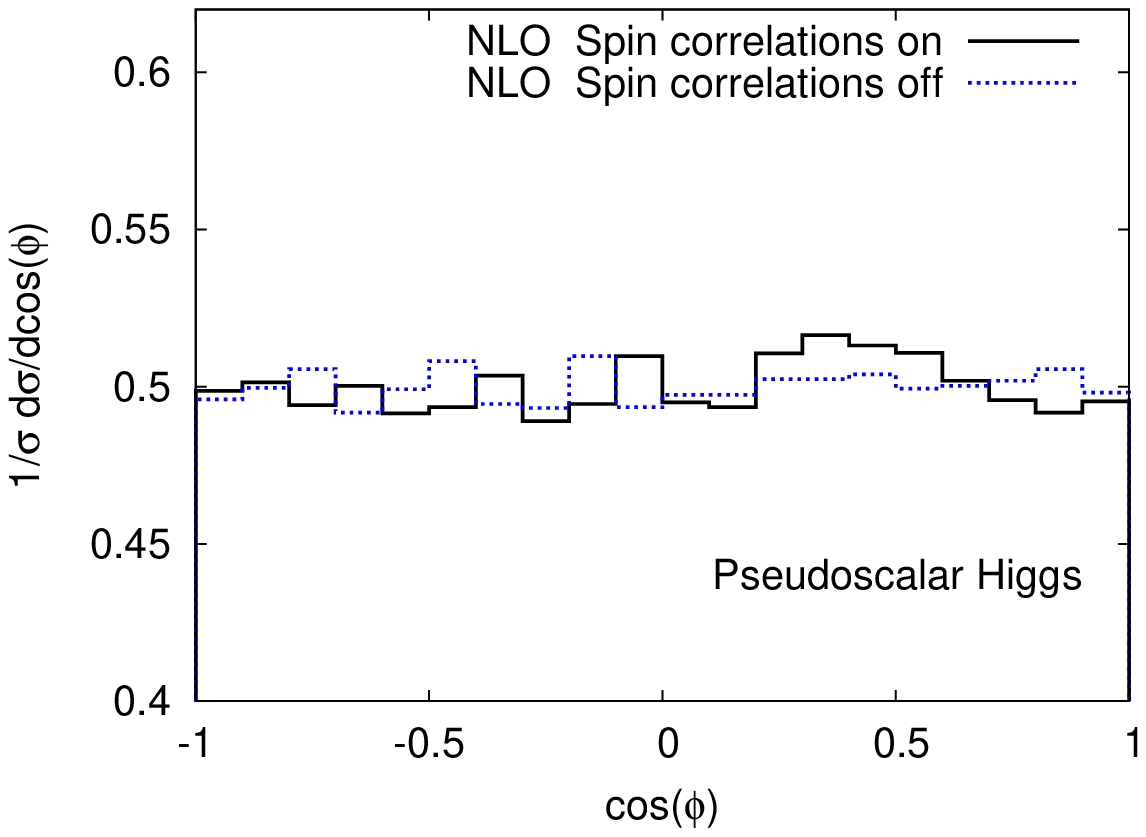}
\caption{Next-to-leading-order cross sections differential in
  $p_T(l^+)$ (left pane) and in $\cos \phi$ (right pane) for $t
  \bar t A$ events with or without spin correlation effects.  Events
  were generated with \amcatnlob, then decayed with \madspinb, and
  finally passed to \herwig for shower and hadronization.  }
\label{fig:Att_distr}
\end{figure}

The results for the pseudo-scalar Higgs boson are shown in
Figure~\ref{fig:Att_distr}. The effects of the spin correlations on
the transverse momentum of the charged lepton are similar as in the
case of a scalar Higgs boson: about 10\% at small $p_T$, increasing to
about 40\% at $p_T=200$ GeV. On the other hand, the $\cos(\phi)$ does
not show any significant effect from the spin-correlations. Therefore
this observable could possibly help in determining the CP nature of
the Higgs boson, underlining the importance of the inclusion of the
spin correlation effects.

\section{Conclusion}

In this paper, we discussed the decay of  events in a LHE file.
We showed that the decay can be treated in a completely 
generic way in the \madgraph  framework.
The procedure is particularly efficient to 
generate unweighted decayed events, while preserving 
spin correlation effects at tree-level accuracy.
Some features associated with the finite width of the 
resonances to be decayed are also successfully reproduced.
The practical implementation of this procedure
--dubbed \madspinb-- 
makes use of the user-friendly \madgraph  interface, 
so that it can be trivially applied to a very large category 
of processes. 
As an illustration we applied the tool to the case of top quark 
pair production in association with a Higgs boson at the LHC, 
and showed for the first time predictions at next-to-leading order 
including spin correlation effects in the angular distributions 
of the leptons.

Limitations of the approach are inherent to restrictions on the validity of the
narrow width approximation. As an example, distributions of events
very close to threshold are sensitive to finite width effects,
 and cannot be predicted accurately within
the procedure discussed in this paper.
Practical limitations are also expected depending  on the complexity
of the process at work. In particular,
the unweighting procedure in  \madspin has not been optimized 
for decay chains which cannot be
split into a sequence of two-body decays. This could be improved by
boosting the efficiency of the unweighting procedure  with the use of
adaptive Monte Carlo techniques.

Within the range of validity of the method, \madspin 
gives access to a large set of applications. 
In particular, it should be stressed that
the algorithm also applies to any BSM decay chain processes 
that can be handled in the narrow width approximation.
The most obvious benefit of this work is that \madspin provides
a very convenient way  to decay events generated at next-to-leading-order 
accuracy, as it is naturally embedded in a Monte Carlo 
scheme. Although only tree-level matrix elements 
are used to unweight the decay configurations, 
we have shown for specific processes that 
this procedure captures essentially all spin correlation effects 
as predicted by a full next-to-leading-order calculation.
A systematic study of the level of accuracy delivered by this 
procedure for generic processes is beyond the scope of this paper.
Nevertheless, a plausible assumption is that information from 
one-loop corrections in the production process are irrelevant 
as far as spin correlations are concerned.
Therefore \madspin --in combination with a next-to-leading-order
Monte Carlo generator-- is expected to give  access 
to event generation with an improved accuracy in many instances.

\acknowledgments
We would like to thank Stefano Frixione, Eric Laenen and Fabio Maltoni 
for many useful discussions on this project and for their 
comments on the manuscript.
P.A. is supported by a Marie Curie Intra-European Fellowship 
(PIEF-GA-2011-299999 PROBE4TeVSCALE).
O.M. is a  fellow of the Belgian American Education Foundation. His work is partially supported by the IISN ``MadGraph'' convention 4.4511.10.
The work of R.R. is supported in part by ``Stichting voor Fundamenteel Onderzoek der Materie (FOM)'', 
which is financially supported by the ``Nederlandse organisatie voor Wetenschappelijke Onderzoek (NWO)''

\appendix

\section{Manual}

The \madspin program is part of the \madgraph distribution from version 2.0.0
onwards.
As a consequence, the \madgraph website 
can be used for downloading the code but also to submit questions or report bugs.
The code can either be used in standalone mode or called by another program of the \madgraph Suite (\ie \madevent or \amcatnlob).
In standalone, one launches a \madspin session by typing:
\begin{verbatim}
./MadSpin/madspin
\end{verbatim}
This opens a prompt, similar to that of the \madgraph Suite. The following commands are then available:

\begin{itemize}
\item {\bf import FILE:} imports the LO or NLO lhe file and reads the associated header in order to load relevant information, \eg the relevant model.
\item {\bf define LABEL = PART1 PART2 ...  :} allows to define an additional multi-particle tag. The multi-particle tags defined in the header of the file are automatically recognized.
\item {\bf decay PROCESS:} specifies the decay branch initiated by an unstable  particle to be decayed by \madspinb. Multi-particle tags can be used for the final state particles.
\item {\bf set OPTION VALUE:} allows to change some internal options of \madspin such as the seed or the value of the maximal weight in the unweighting procedure. 
Type ``help options'' in the interface for more details on the various available options.
\item {\bf launch:} runs \madspin according to the specified options/decay channels.
\item {\bf help COMMAND:} provides detailed information on a specific command. 
\end{itemize}

As stated above, \madspin can also be called directly via the \madeventb/\amcatnlo script (via the launch command or the ./bin/generate\_events script). 
In this case, the script will request some information about the  programs to run. 
The exact question depends on the programs installed on your computer and on the QCD order of the process (LO or NLO),
but it typically looks like this:

\begin{verbatim}
Which programs do you want to run?
  0 / auto    : running existing card
  1 / parton  :  Madevent
  2 / pythia  : MadEvent + Pythia.
  3 / pgs     : MadEvent + Pythia + PGS.
+10 / +madspin: adding MadSpin to the run (before running Pythia) 
[0, auto, 1, parton, 11, parton+madspin, 2, pythia, 12, ... ]
\end{verbatim}

If you choose to run \madspin (\ie if you enter one of the following answers: 11, 12, 13, parton+madspin, pythia+madspin, pgs+madspin),
the program will propose, in a second step, to edit the madspin\_card.dat.
This file contains the command lines associated with the \madspin program
that we just described.
Note that \madgraph will figure out which event file has to be decayed, so that you do not need to include the ``import'' command in this file.
After these questions have been answered, \madgraph proceeds with the 
Monte Carlo generation: (1) it generates the LHE event file before decay,
(2) it executes \madspin and (3) if requested, it submits the decayed events
to a shower/hadronisation program.

\end{document}